\documentclass[10pt,osajnl,amssymb,preprint,showpacs,twocolumn]{revtex4}
\usepackage{hyperref}
\usepackage{amsmath}
\usepackage{graphicx}

\begin{document}


\title{Semi-discrete composite solitons in arrays of quadratically nonlinear waveguides}
\author{Nicolae C. Panoiu and Richard M. Osgood, Jr.}
\affiliation{Department of Applied Physics and Applied Mathematics, Columbia University, New York,
New York 10027}

\author{Boris A. Malomed}
\affiliation{Department of Interdisciplinary Studies, Faculty of Engineering, Tel-Aviv University,
Tel-Aviv 69978, Israel}

\begin{abstract}
We demonstrate that an array of discrete waveguides on a slab substrate, both featuring the $\chi
^{2}$ nonlinearity, supports stable solitons composed of discrete and continuous components. Two
classes of fundamental composite solitons are identified: ones consisting of a discrete
fundamental-frequency (FF) component in the waveguide array, coupled to a continuous
second-harmonic (SH) component in the slab waveguide, and solitons with an inverted FF/SH
structure. Twisted bound states of the fundamental solitons are found too. In contrast with usual
systems, the \emph{intersite-centered} fundamental solitons and bound states with the twisted
continuous components are stable, in an almost entire domain of their existence.
\end{abstract}


\maketitle

Quadratically nonlinear ($\chi ^{(2)}$) media, continuous or discrete, provide favorable
conditions for the creation of optical solitons. The wave-vector mismatch and $\chi ^{(2)}$
coefficient are efficient control parameters in this context, and the solitons display a variety
of features due to their ``multicolor" character. Accordingly, a great effort has been invested in
the study of solitons in continuous \cite{hk93prl,twh95prl} (as reviewed in
\cite{review1,review2}) and (quasi-)discrete
\cite{bcc97pre,ppl98pre,ktv04ol,kvt04ol,iss04prl,mkf02pre,cls03n} $\chi ^{(2)}$ media. In both
cases, the solitons can find applications to all-optical switching \cite{ct99ol,ppl03ol,cls03n}
and light-beam steering.\cite{sbs98oqe,vmk04pre}

Waveguide arrays, i.e., one-dimensional (1D) discrete systems, exhibit properties that are absent
in continuous media, such as anomalous or managed diffraction.\cite{cls03n} Accordingly, discrete
solitons are drastically different from their counterparts in continuous media, as was first
predicted in the context of the $\chi ^{(3)}$ nonlinearity \cite{cj88ol}. Here we propose
\textit{semi-discrete} composite\emph{\ }solitons in $\chi ^{(2)}$ optical systems, that contain
\emph{both} discrete and continuous components, each carrying either the fundamental-frequency
(FF) or second-harmonic (SH) wave. We demonstrate that \emph{stable} semi-discrete solitons can be
readily formed in a waveguide array coupled to a slab waveguide, both structures being made of a
quadratically nonlinear material. We study two most interesting species of semi-discrete solitons.
\textit{Type-A} ones consist of a discrete FF component in the waveguide array, coupled to a
continuous SH component in the slab waveguide. Conversely, \textit{type-B} solitons feature
continuous FF and discrete SH components in the slab and discrete array, respectively.

\begin{figure}[b]
\centering \leavevmode
\includegraphics[width=2.5cm, angle=270]{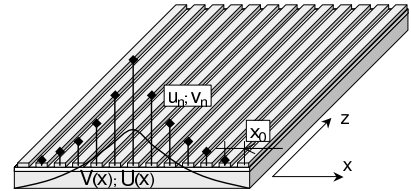}
\caption{Schematic of the proposed setting.} \label{fcrystal}
\end{figure}

The proposed setting is displayed in Fig. \ref{fcrystal}. It includes the periodic array of
waveguides, with a spacing $x_{0}$, mounted on top of (or buried into) the slab waveguide. Both
the array and slab are made of a $\chi ^{(2)}$ material, such as $\mathrm{LiNbO_{3}}$ or
$\mathrm{KTiOPO_{4}}$. A rigorous coupled-mode theory for such composite waveguides can be
developed by a straightforward generalization of that available for a single discrete waveguide
coupled to the slab \cite{vh96oc}. Thus, we arrive at a system including a set of ordinary
differential equations for the discrete array, coupled to a partial differential equation for the
slab. In case A, the coupled equations take the following normalized form:
\begin{subequations}
\label{caseIphiPsi}
\begin{eqnarray}
&&i\frac{\displaystyle d\phi _{n}}{\displaystyle d\zeta }+\varrho \left( \phi _{n-1}+\phi
_{n+1}\right) +\phi _{n}^{\ast }\Psi (\zeta ,n)=0,
\label{Psi} \\
&&i\frac{\displaystyle\partial \Psi }{\displaystyle\partial \zeta
}+\frac{\displaystyle1}{\displaystyle2}\frac{\displaystyle\partial ^{2}\Psi
}{\displaystyle\partial \eta ^{2}}+\beta \Psi +\frac{\displaystyle1}{\displaystyle2}\sum_{n}\phi
_{n}^{2}\delta (\eta -n)=0,
\end{eqnarray}and in case B they are written as
\end{subequations}
\begin{subequations}
\label{caseIIphiPhi}
\begin{eqnarray}
&&i\frac{\displaystyle\partial \Phi }{\displaystyle\partial \zeta
}+\frac{\displaystyle1}{\displaystyle2}\frac{\displaystyle\partial ^{2}\Phi
}{\displaystyle\partial \eta ^{2}}+\Phi ^{\ast }\sum_{n}\psi _{n}\delta (\eta
-n)=0,  \label{psi} \\
&&i\frac{\displaystyle d\psi _{n}}{\displaystyle d\zeta }+\beta \psi _{n}+\varrho \left( \psi
_{n-1}+\psi _{n+1}\right) +\frac{\displaystyle1}{\displaystyle2}\Phi ^{2}(\zeta ,n)=0.
\end{eqnarray}
Here, the normalized coordinates are $\zeta =z/z_{0}$ and $\eta =x/x_{0}$, where $z$ and $x$ are,
respectively, the distances in the propagation and transverse directions, $z_{0}=Kx_{0}^{2}$, $K$
is the propagation constant of the corresponding continuous mode, $\beta $ is the effective wave
vector mismatch, $\delta (\eta )$ is the delta-function, and $\varrho =c_{d}z_{0}$ is the
normalized coupling constant between adjacent waveguides in the array, where $c_{d}$ is the
coupling constant in physical units, as predicted by the coupled-mode theory.\cite{vh96oc} The
normalized field amplitudes at the FF, $\phi _{n}$ and $\Phi $, and at the SH, $\psi _{n}$ and
$\Psi $, are proportional to their counterparts, $u_{n},U$ and $v_{n},V$, measured in physical
units: $\phi _{n}=(z_{0}/\sqrt{x_{0}})\sqrt{2\gamma _{d}\gamma _{c}}u_{n}$, $\Psi =z_{0}\gamma
_{d}V$ and $\psi _{n}=(z_{0}/x_{0})\gamma _{c}v_{n}$, $\Phi =(z_{0}/\sqrt{x_{0}})\sqrt{2\gamma
_{d}\gamma _{c}}u_{n}U$, where $\gamma _{d,c}=\frac{\epsilon _{0}\omega
}{4P_{d,c}\sqrt{P_{c,d}}}{\int_{A_{d,c}}}dA\mathbf{\hat{e}}_{n,c}^{\ast }(\omega )\cdot
\mathbf{\hat{\chi}}(2\omega ,\omega )\mathbf{\hat{e}}_{n,c}^{\ast }(\omega
)\mathbf{\hat{e}}_{c,n}(2\omega )$, $\gamma _{c,d}=\frac{\epsilon _{0}\omega
}{2P_{d,c}\sqrt{P_{c,d}}}{\int_{A_{c,d}}}dA\mathbf{\hat{e}}_{c,n}^{\ast }(2\omega )\cdot
\mathbf{\hat{\chi}}(\omega ,\omega )\mathbf{\hat{e}}_{n,c}(\omega )\mathbf{\hat{e}}_{n,c}(\omega
)$ in cases A and B, respectively, with $P_{d}$ and $P_{c}$ the power in the stripe waveguide and
the power density in the slab waveguide, $A_{d}$ and $A_{c}$ the transverse areas of the stripe
waveguide and slab waveguide between two adjacent stripes, respectively, and
$\mathbf{\hat{e}}_{n}(x,y)$ and $\mathbf{\hat{e}}_{c}(x)$ fundamental modes of the stripe and slab
waveguides. In this paper, we focus on the case of $\varrho =1$, which adequately represents the
generic situation.

\begin{figure}[t]
\centering \leavevmode
\includegraphics[width=6.5cm]{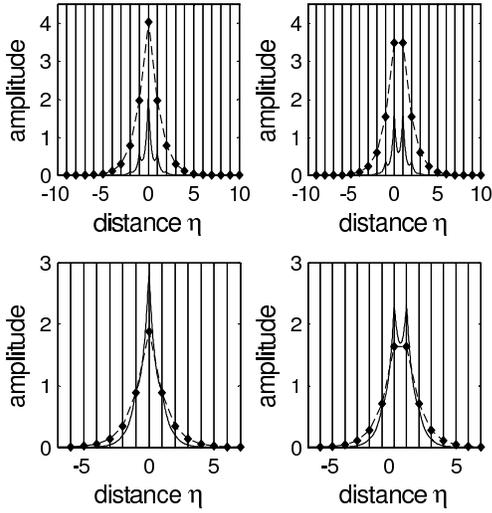}
\caption{Field profiles of composite solitons. Here and in the next figure, top and bottom panels
correspond to the solitons of types A and B, respectively, whereas left and right panels display
odd and even solitons. Vertical lines designate the location of discrete waveguides. The wave
vector mismatch is $\protect\beta =0$, and the soliton wavenumber is $\protect\lambda =3$ and
$\protect\lambda =1.5$, in cases A and B, respectively.} \label{foddeven}
\end{figure}

The above equations were derived for the case when the FF and SH fields have the same (TE or TM)
polarization. Using the general coupled-mode description \cite{vh93jlt}, the system can be
extended for the case of two polarizations, which will lead to a type-II $\chi ^{(2)}$
interaction\cite{review1,review2} involving two different FF components. This case will be
presented elsewhere.

Both variants of the model (A and B) neglect direct FF-SH interactions in the same waveguide, as
they are suppressed by the large natural mismatch, while we assume that care is taken to minimize
the mode mismatch, between the continuous and discrete waveguides. Such requirements can be easily
fulfilled, as the geometry gives rise to different propagation constants for the same frequency in
the waveguides of the two types. A model which takes into regard the residual SH-FF coupling in
each waveguide can be easily considered too, but cases A and B, as defined above, are the most
interesting ones. Note also that in this geometry the overlap between the FF and SH is smaller, as
compared to the case in which FF-SH interactions in the same waveguide are employed.

\begin{figure}[t]
\centering \leavevmode
\includegraphics[width=6.5cm]{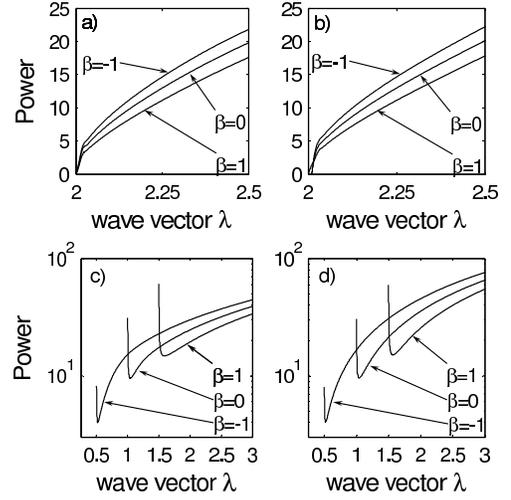}
\caption{The total power $P$ of the semi-discrete solitons \textit{vs.} the wavenumber
$\protect\lambda$.} \label{fpower}
\end{figure}

Composite solitons amount to stationary solutions of systems (\ref{caseIphiPsi}) and
(\ref{caseIIphiPhi}) in the form of $\phi _{n}(\zeta )=\bar{\phi}_{n}\exp (i\lambda \zeta )$,
$\Psi (\eta ,\zeta )=\bar{\Psi}(\eta )\exp (2i\lambda \zeta )$ and $\psi _{n}(\zeta
)=\bar{\psi}_{n}\exp (2i\lambda \zeta )$, $\Phi (\eta ,\zeta )=\bar{\Phi}(\eta )\exp (i\lambda
\zeta )$, respectively, where $\lambda $ is the soliton's wavenumber. Inserting these expressions
in the underlying equations, we solved the resulting systems by dint of the Newton-Raphson method.
Similar to the ordinary discrete solitons, the composite ones can be odd or even: the former ones
are centered at a site of a discrete waveguide, whereas even solitons are intersite-centered.
Typical examples of odd and even composite solitons are displayed for both cases, A and B, in Fig.
\ref{foddeven}.

To examine the stability of the composite solitons, we have first applied the
\textit{Vakhitov-Kolokolov} (VK) criterion \cite{vk73rqe}, which predicts that the necessary
stability condition is $dP/d\lambda >0$, where $P $ is the total power [which is the single
dynamical invariant of Eqs. (\ref{caseIphiPsi}) and (\ref{caseIIphiPhi})], $P(\lambda
)=2\int_{-\infty }^{+\infty }|\left\{ \Psi ,\Phi \right\} |^{2}d\eta +\sum_{n}|\left\{ \phi
_{n},\psi _{n}\right\} |^{2}$, in cases A and B, respectively. The result is displayed in Fig.
\ref{fpower}, which shows that (as may be expected) the solitons exist above the band of linear
waves of the discrete subsystem, i.e., for $\lambda >2$ in case A, and $\lambda >1+\beta /2$ in
case B, and both odd and even solitons are VK-stable in most of their existence domain. The
prediction is significant, as even solitons are \emph{always} unstable in ordinary discrete
systems. Note that the power of odd solitons is slightly smaller than that of even ones.

A noteworthy feature is that A-type solitons exist up to $P=0$, while their B-type counterparts
have a cutoff in terms of $P$. This can be explained by noting that the limit of $P\rightarrow 0$
corresponds to the limit of broad (quasi-continuous) small-amplitude solitons. Straightforward
analysis of Eqs. (\ref{Psi}) demonstrates that this limit exists indeed, reducing to
$i\tilde{\phi}_{\zeta }+\tilde{\phi}_{\eta \eta }+|\tilde{\phi}|^{2}\tilde{\phi}/\left( 2\left(
4-\beta \right) \right) =0$, $\tilde{\Psi}=\tilde{\phi}^{2}/\left( 2(4-\beta )\right) $, with
$\{\tilde{\phi},\tilde{\Psi}\}=\{\phi _{n}e^{-2i\zeta },\Psi e^{-4i\zeta }\}$. On the other hand,
the same limit, if applied to Eqs. (\ref{psi}), amounts to a \emph{linear} asymptotic equation,
$i\Phi _{\zeta }+\Phi _{\eta \eta }=0$, which gives rise to no soliton solutions.

Full stability of the solitons was examined in direct simulations of Eqs. (\ref{caseIphiPsi}) and
(\ref{caseIIphiPhi}), which \emph{completely corroborate} the predictions of the VK criterion. In
particular, those composite solitons of the B-type which are VK-unstable as per Fig. \ref{fpower}
decay into linear waves.

We have also studied the \textit{twisted solitons}, built as out-of-phase bound states of two odd
ones, see examples of A- and B-type solitons with a twisted FF component in Fig. \ref{ftwist}.
Their stability was also examined by means of both the VK criterion (see Fig. \ref{ftwist}) and in
direct simulations. The results demonstrate that the semi-discrete twisted solitons exist if their
wavenumber $\lambda $ exceeds a cut-off value, which depends on the mismatch $\beta $, and they
are \emph{stable} in almost the entire existence domain. By contrast, in continuous $\chi ^{(2)}$
media twisted solitons are \emph{always unstable} \cite{review1,review2}. In a small region near
the cut-off the twisted composite solitons are unstable too. The cut-off being well separated from
the band of linear waves, in simulations the unstable twisted solitons do no decay into linear
waves, but rather evolve into a fundamental odd soliton or split in two such solitons.

\begin{figure}[t]
\centering \leavevmode
\includegraphics[width=6.5cm]{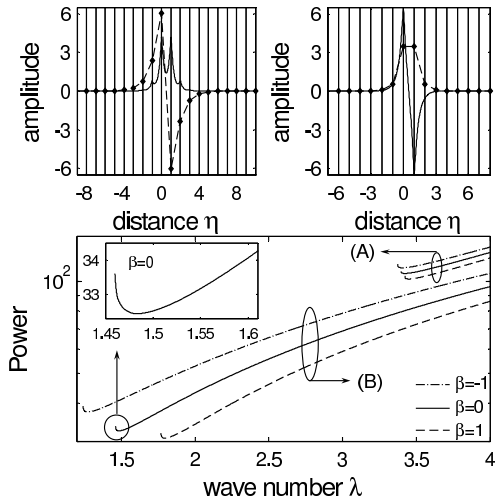}
\caption{Top panels: field profiles of twisted composite solitons of types A (left) and B (right).
In both cases, the wave vector mismatch and soliton's wavenumber are $\protect\beta =0$ and
$\protect\lambda =3.6$. The bottom panel shows the total power $P$ \textit{vs.} $\protect\lambda
$, the inset blowing up a vicinity of the cutoff point (stability border) on the curve for
$\protect\beta =0$.} \label{ftwist}
\end{figure}

To summarize, we have shown that stable composite solitons, fundamental and twisted ones, can be
supported by a slab waveguide coupled to an array of waveguides, both made of a $\chi ^{(2)}$
material. This new species of \textit{semi-discrete solitons} features unusual properties,
\textit{viz}., stability of intersite-centered fundamental solitons, and of states with the
twisted continuous component. One can envisage that this system may also support
\textit{walking}\cite{tmm96prl} semi-discrete solitons, and that similar solitons can be found in
dual discrete-continuous systems with the Kerr nonlinearity, as well as in multidimensional $\chi
^{(2)}$ systems.

This work has been supported by the NSF Grant No. ECS-0523386 and DoD STTR Grant No.
FA9550-04-C-0022.

\end{subequations}

\end{document}